\documentclass[twocolumn,floatfix,superscriptaddress,a4paper,showkeys,nofootinbib,reprint]{revtex4-1}
\usepackage[colorlinks=true,linktocpage=true,linkcolor=blue,citecolor=blue,allcolors=blue]{hyperref}
\usepackage{epsfig}
\usepackage{latexsym}
\usepackage[utf8]{inputenc}
\usepackage{xspace}
\usepackage{indentfirst}
\usepackage{enumitem}
\usepackage{color}
\usepackage{setspace}
\usepackage{lipsum}
\usepackage{graphicx}
\usepackage{todonotes}
\usepackage{multirow}
\usepackage{amsmath}
\usepackage{amssymb}
\usepackage[english]{babel}
\usepackage{url}

\newcommand{\mean}[1]{\langle #1 \rangle}

\newcommand{\eq}[1]{\begin{align} #1 \end{align}}

\global\long\def\ave#1{\left\langle #1 \right\rangle }%

\UseRawInputEncoding
\begin{document}
	
\title{Critical point particle number fluctuations from molecular dynamics}
\author{Volodymyr A. Kuznietsov}
\affiliation{Bogolyubov Institute for Theoretical Physics, 03680 Kyiv, Ukraine}
\affiliation{Physics Department, Taras Shevchenko National University of Kyiv, 03022 Kyiv, Ukraine}
\author{Oleh Savchuk}
\affiliation{Bogolyubov Institute for Theoretical Physics, 03680 Kyiv, Ukraine}
\affiliation{Frankfurt Institute for Advanced Studies, Giersch Science Center, Ruth-Moufang-Str. 1, D-60438 Frankfurt am Main, Germany}
\affiliation{
GSI Helmholtzzentrum f\"ur Schwerionenforschung GmbH, Planckstr. 1, D-64291 Darmstadt, Germany}

\author{Mark I. Gorenstein}
\affiliation{Bogolyubov Institute for Theoretical Physics, 03680 Kyiv, Ukraine}
\affiliation{Frankfurt Institute for Advanced Studies, Giersch Science Center, Ruth-Moufang-Str. 1, D-60438 Frankfurt am Main, Germany}
\author{Volker Koch}
\affiliation{Nuclear Science Division, Lawrence Berkeley National Laboratory, 1 Cyclotron Road,  Berkeley, CA 94720, USA}
\author{Volodymyr Vovchenko}
\affiliation{Nuclear Science Division, Lawrence Berkeley National Laboratory, 1 Cyclotron Road,  Berkeley, CA 94720, USA}

\date{\today}
\begin{abstract}
	We study fluctuations of particle number in the presence of critical point by utilizing molecular dynamics simulations of the classical Lennard-Jones fluid in a periodic box.
	The numerical solution of the $N$-body problem naturally incorporates all correlations, exact conservation laws, and finite size effects, allowing us to study the fluctuation signatures of the critical point in a dynamical setup.
	We find that large fluctuations associated with the critical point are observed when measurements are performed in coordinate subspace, but, in the absence of collective flow and expansion, are essentially washed out when momentum cuts are imposed instead.
	We put our findings in the context of event-by-event fluctuations in heavy-ion collisions.
\end{abstract}

\maketitle
\section{Introduction}

A critical point~(CP) is the endpoint of a first-order phase transition line where the phase boundaries vanish.
It is a ubiquitous phenomenon which occurs in many different physical systems, including most atomic and molecular systems, ferromagnets, cold nuclear matter, and potentially hot QCD matter.
A generic feature of the CP is growth of the thermal fluctuations in its vicinity, which, for an infinite system, become divergent at the CP.
For instance, large and long range density fluctuations near the CP of a liquid-gas transition explain the well known phenomenon of critical opalescence.

Theoretically, the thermal fluctuations of a (conserved) particle number are encoded in the equation of state and can be most easily characterized within grand-canonical statistical mechanics. The (scaled) variance of particle number fluctuations in the grand-canonical ensemble reads~\cite{huang1987statistical}
\eq{
\frac{\mean{\Delta N^2}}{\mean{N}} = \frac{T}{\left(\frac{\partial{\tilde{p}}}{\partial \tilde{n}}\right)_{\tilde{T}}}~.
}
At the CP one has $\left(\frac{\partial{\tilde{p}}}{\partial \tilde{n}}\right)_{\tilde{T}} = 0$, thus the fluctuations formally diverge.

Using event-by-event fluctuations is the key idea in the experimental search for the QCD CP at finite baryon density with heavy-ion collisions~\cite{Stephanov:1999zu,Bzdak:2019pkr}. Here the baryon current plays the role of the conserved particle number and the presence of the QCD CP should manifest itself in the enhanced fluctuations of proton number~\cite{Hatta:2003wn}, as well as possibly nonmonotonic collision energy dependence of the high-order measures like skewness and kurtosis~\cite{Stephanov:2008qz,Stephanov:2011pb}.
The corresponding measurements have been performed by different experiments like STAR~\cite{STAR:2020tga,STAR:2021iop}, HADES~\cite{HADES:2020wpc}, and ALICE~\cite{ALICE:2019nbs}.
A definitive interpretation of these measurements is still elusive, but, coupled with the available constraints from first-principle lattice QCD simulations at small baryon densities~\cite{Bazavov:2017dus,Vovchenko:2017gkg,Borsanyi:2020fev}, there are indications that the QCD critical region can likely only be created in heavy-ion collisions at sufficiently large baryon densities, corresponding to collision energies of $\sqrt{s_{\rm NN}} \lesssim 7.7$~GeV.
At these collision energies the production of antibaryons can be neglected and the analysis amounts to the study of the event-by-event distribution of proton number.

Theoretical interpretation of experimental results on fluctuations is challenging because grand-canonical statistical mechanics is not directly applicable to the conditions realized in the experiment~\cite{Koch:2008ia,Vovchenko:2021gas}. In particular, the growth of critical fluctuations is restricted by both the finite sizes and lifetimes of the systems created in the experiment~\cite{Berdnikov:1999ph,Poberezhnyuk:2020ayn} as well as the exact global conservation of baryon number~\cite{Bzdak:2012an}.
Furthermore, measurements are necessarily performed in momentum space whereas the physics of the CP and its associated correlations is usually discussed in
configuration space. 
Methods to correct fluctuation measurements for global conservation laws have been recently developed~\cite{Vovchenko:2020tsr,Vovchenko:2020gne,Vovchenko:2021yen} but their limits of applicability near the CP and at lower collision energies remain unclear.
A quantitative framework for critical fluctuations in heavy-ion collisions based on fluctuating hydrodynamics is under development~\cite{Bluhm:2020mpc,An:2021wof}.

In the present work we study critical fluctuations of (conserved) particle number within molecular dynamics~(MD) simulations of the Lennard-Jones~(LJ) fluid.
The LJ fluid corresponds to a system of non-relativistic particles with attractive and repulsive interactions, which contains a first-order phase transition and the associated CP.
This system is quite different from the hot QCD matter near the QCD CP, where both the hadronic and partonic degrees of freedom are relevant.
Nevertheless, MD simulations of the LJ fluid provide a microscopic approach to fluctuations near the CP, and allow to study deviations from the baselines predicted by the grand-canonical statistical mechanics.
They also contain all (classical) correlations in the system, i.e. no approximations like mean-field based description are applied.
In particular, using MD one can obtain particle number distributions affected by the CP which could well mimick the event-by-event distributions of protons near the QCD CP. 
Previously, the LJ model has been used in various studies of nuclear matter and QCD~\cite{Dorso:1988doe,Lai:2009cn,Moretto_2011,Shuryak:2018lgd}.
Here we perform MD simulations of the LJ fluid in a box with periodic boundary conditions at both near and away from the CP.
We study particle number fluctuations inside a coordinate space subvolume and how these fluctuations relate to the grand-canonical susceptibilities, with a focus on the role of finite system size and global conservation laws.
We then analyze how the behavior of fluctuations changes when the analysis is performed in momentum rather than coordinate space.
The studies in the present work are restricted to box simulations but can be extended in future works to describe expanding systems, reflecting better the conditions realized in heavy-ion experiments.

The paper is organized as follows.
The LJ fluid as well as the main quantities of study are introduced in Sec.~\ref{sec:LJ}.
The details of the MD simulations are described in Sec.~\ref{sec:MD} and the results presented in Sec.~\ref{sec:results}.
The summary and outlook in Sec.~\ref{sec:summary} closes the article.

\section{Lennard-Jones fluid}
\label{sec:LJ}

The LJ fluid is a system of particles interacting via the LJ potential:
\eq{V_{\rm LJ}(r) = 4\varepsilon\left[\left(\frac{\sigma}{r}\right)^{12} - \left(\frac{\sigma}{r}\right)^{6}\right].}
Here the first term corresponds to the repulsive core at short distances whereas the second term describes the attraction at an intermediate range.
The two parameters -- $\sigma$ and $\varepsilon$ -- define the size of the repulsive core and the depth of the attractive well, respectively.
It is customary to treat $\sigma$ and $\varepsilon$ as length and energy scales and work with dimensionless variables.
In this case the reduced potential $\tilde{V}_{\rm LJ} = V_{\rm LJ} / \varepsilon$  reads
\eq{
\tilde{V}_{\rm LJ}(\tilde{r}) = 4\left( \tilde{r}^{-12} - \tilde{r}^{-6} \right),
}
with $\tilde{r} = r/\sigma$ being the reduced distance.
The reduced thermodynamic variables are the temperature $\tilde{T} = T/(k_B \varepsilon)$, particle number density $\tilde{n} = n \sigma^3$, and pressure $\tilde{p}=p\sigma^3/\varepsilon$. 
The particle's mass can be utilized to define the dimensionless time variable, $\tilde{t} = t \sqrt{\varepsilon /(m\sigma^2)}$.

The Lennard-Jones fluid possesses a rich phase diagram, with phase transitions between various gas, liquid, and solid phases~(see e.g.~\cite{STEPHAN2020112772} for an overview).
The CP of a liquid-gas transition is of primary interest in the present work.
The CP location has been estimated from numerous MD simulations, yielding~\cite{doi:10.1021/acs.jcim.9b00620}
\eq{
\label{eq:TcLJ} 
\tilde{T}_c &  = 1.321 \pm 0.007~ , ~~~
\tilde{n}_c  = 0.316 \pm 0.005~.
}
The CP is characterized by the critical pressure of $\tilde{p}_c = 0.129 \pm 0.005$.
This gives the compressibility factor $Z \equiv \tilde{p}/(\tilde{n}\tilde{T})$ at the CP of $Z_c 
\approx 0.309$.
For comparison, the CP compressibility factor in the van der Waals model is $Z_c^{\rm vdW} = 3/8 = 0.375$, i.e. about 20\% higher.

\subsection{Particle number fluctuations}

Thermal fluctuations are expected to be large near the CP.
In particular, the macroscopic growth of particle density fluctuations leads to the phenomenon of critical opalescence.
Formally, these fluctuations can be analyzed in the framework of the grand-canonical statistical mechanics, which corresponds to a system in contact with the heat bath with which it can exchange particles.
The variance of particle number is given by the derivative of the mean particle number with respect to the chemical potential,
\eq{
\mean{\Delta N^2} \equiv \mean{(N-\mean {N})^2} = T \, \left(\frac{\partial{\mean{N}}}{\partial \mu}\right)_{T,V}~,
}
where the symbol $\langle \ldots \rangle$ denotes the grand-canonical averaging.
Using the thermodynamic identity $\left(\frac{\partial{\mean{N}}}{\partial \mu}\right)_{T,V} = \mean{N} / \left(\frac{\partial{p}}{\partial n}\right)_{T}$, one can express the fluctuations in terms of the derivative of pressure with respect to density.
It is instructive to consider the so-called scaled variance $\omega = \mean{\Delta N^2} / \mean{N}$ which is an intensive measure of fluctuations:
\eq{\label{eq:w}
\omega 
 = \frac{T}{ \left(\frac{\partial{p}}{\partial n}\right)_{T}} = \frac{\tilde{T}}{ \left(\frac{\partial{\tilde{p}}}{\partial \tilde{n}}\right)_{\tilde{T}}}
 = \left[ Z + \tilde{n} \left( \frac{\partial Z}{\partial \tilde{n}} \right)_{\tilde{T}} \right]^{-1}~.
}
The scaled variance grows in the vicinity of the CP and diverges at the CP where $(\partial \tilde{p} / \partial \tilde{n})_{\tilde{T}} = 0$.

\subsection{Virial expansion}

The equation of state of LJ fluid in a closed form is not known.
In the low-density limit, however, it can be approximated using the virial expansion.
The virial expansion for the compressibility factor reads
\eq{\label{eq:Pvir}
Z \equiv \frac{\tilde{p}}{\tilde{n}\tilde{T}} = 1 + \sum_{k = 2}^{\infty} \tilde{B}_k \, (\tilde{T}) \, \tilde{n}^{k-1}.
}
Here $\tilde{B}_k \equiv B_k \, \sigma^{-3k}$ are the (reduced) virial coefficients.
The leading coefficient $\tilde B_2(\tilde T)$ can be calculated analytically~\cite{VARGAS200192}, while for the higher-order ones high-precision numerical data in broad temperature range are available~\cite{shaul2010effect,schultz2009sixth}.
Using the virial expansion~\eqref{eq:Pvir} one can rewrite the scaled variance as follows
\eq{\label{eq:wvir}
\omega(\tilde{T}, \tilde{n}) = \frac{\mean{\Delta N^2}}{\mean{N}} = \left[1 +  \sum_{k = 2}^{\infty} k \, \tilde{B}_k \, (\tilde{T}) \, \tilde{n}^{k-1} \right]^{-1}~.
}

The temperature derivatives of the virial coefficeints can be utilized to calculate the energy per particle at given $\tilde{T}$ and $\tilde{n}$~(see the details in Appendix~\ref{app:virial}):
\eq{\label{eq:uvir}
\frac{\tilde{U}}{N} = \frac{3}{2} \tilde{T} - \sum_{k=2}^{\infty}  \frac{\tilde{T}^2 \, \tilde{B}'_k(\tilde{T})}{k-1} \tilde{n}^{k-1}~.
}

A truncated virial expansion gives a good approximation of the equation of state at sufficiently low densities where it  converges rapidly. We will utilize the virial expansion in Eqs.~\eqref{eq:Pvir} and~\eqref{eq:wvir} to test the accuracy of our MD simulations in those regions where the virial expansion is applicable.
The parameterizations for the virial expansion coefficients are given in Appendix~\ref{app:virial}.

\section{Molecular dynamics simulation}
\label{sec:MD}

MD simulations are performed by numerically integrating the Newton's equations of motion using the Velocity Verlet integration method.
The simulations are done for a system of $N$ particles with periodic boundary conditions in the minimum-image convention form.\footnote{In the minimum image convention form each particle interacts only with the nearest images of all other particles across the simulation cube and its neighboring periodic images, see e.g.~\cite{allen2017computer} for details.}
Periodic boundary conditions is the most common choice in molecular dynamics simulations.
Other boundary conditions, such as reflecting walls, are also possible, and may lead to quantitative differences of the system-size dependence of the results.
The integration time step is $\Delta \tilde{t} = 0.004$ by default, and where necessary, in particular at high densities, it is reduced to a smaller value to ensure the stability of the numerical integration.

The calculations are performed for fixed values of the particle number $N$ and density $\tilde{n}$.
The desired particle number density $\tilde{n}$ is achieved by choosing appropriately the length $\tilde{L}$ of the cubic simulation box, namely $\tilde{L} = (N/\tilde{n})^{1/3}$.
The simulations are carried out either in the microcanonical ensemble, where the total energy $\tilde{U}$ is fixed, or in a canonical-like ensemble that keeps the kinetic temperature constant through an additional constraint in the equations of motion~(see Sec. 3.8.2 in \cite{allen2017computer}).

The initial state is prepared by distributing the particle coordinates over a regular cubic lattice and sampling their velocities from the Maxwell-Boltzmann distribution corresponding to the desired temperature $\tilde{T}$.
Then, the velocity components of all the particles are shifted such that the total momentum in the system is zero.
Finally, all the velocities are rescaled by a factor such that the total system energy matches the desired total energy~(microcanonical ensemble) or the total kinetic energy matches the one given by the desired system temperature~(canonical-like ensemble).

The MD simulation is split into two stages: the equilibration and production.
During the equilibration stage the system evolves toward thermodynamic equilibrium.
We assign an equilibration time of $\tilde{t}_{\rm eq} = 50$ for the duration of this stage to ensure that equilibrium is achieved.
This has been checked by observing the behavior of the kinetic temperature~(or the mean energy per particle if the canonical-like ensemble is employed), which, once equilibrium is achieved, exhibits small fluctuations around the true temperature as function of time. 

The production stage begins at $\tilde{t} = \tilde{t}_{\rm eq}$ and is simulated for a time interval of $\tilde \tau$. All of the observables of interest are calculated as time averages during the production phase, i.e. a quantity $A$ which at any given time moment is a function of phase space coordinates $\{\mathbf{\tilde{r}}_i,\mathbf{\tilde{v}}_i\}$ is calculated as
\eq{\label{eq:ergod}
\mean{A} = \frac{1}{\tilde{\tau}} \int_{\tilde{t}_{\rm eq}}^{\tilde{t}_{\rm eq} + \tilde{\tau}} \, A(\{\mathbf{\tilde{r}}_i(\tilde{t}),\mathbf{\tilde{v}}_i(\tilde{t})\}) d \tilde{t}~.
}
In accordance with the ergodic hypothesis, in the limit $\tilde{\tau} \to \infty$ the time average $\mean{A}$ reduces to the ensemble average $\overline{A}$, thus MD simulations over a sufficiently long period of time give access to various statistical mechanics properties of the LJ fluid.

In practice, the integral in Eq.~\eqref{eq:ergod} is evaluated as an average of all the observations taken after each numerical integration time step during the MD simulation.
Furthermore, when the value of $\tilde \tau$ is finite, as is the case in any MD simulation, the expression $\eqref{eq:ergod}$ approximates $\mean{A}$ with a non-vanishing statistical error.
Extra care should be taken to estimate this error correctly, as the consecutive samples taken from the MD simulation unavoidably exhibit autocorrelations. Here we follow the procedure described in~\cite{allen2017computer} to estimate the statistical errors of all our calculations of fluctuations that are corrected for the noise from autocorrelations, while the bootstrap method is used to calculate errors for other observables.

Equation~\eqref{eq:ergod} can be used to calculate a variety of quantities.
For instance, the total energy $\tilde{U}$ is calculated straightforwardly and can be used to determine the energy per particle $\tilde{u} = \tilde{U} / N$ at a given $\tilde{T}$ and $\tilde{n}$ by utilizing the canonical-like ensemble simulations\footnote{In the microcanonical ensemble, where the energy is fixed, monitoring $\tilde{u}$ can be used to check the numerical stability and accuracy of simulations.}.
The temperature $\tilde{T}$ corresponds to the average kinetic energy in the system, thus it can be determined through the time average of the mean velocity squared, i.e. $\tilde{T} = \mean{\mathbf{\tilde{v}}^2} / 3$. 
Finally, the instantaneous pressure can be determined through the virial theorem.
It is calculated as the average over the diagonal components of the negative (non-relativistic) stress tensor~\cite{irgens2008continuum}.
The expression that is appropriate for use in MD simulations with periodic boundary conditions utilizing the minimum image convention reads~\cite{allen2017computer}:
\eq{\label{eq:pirtheo}
\tilde{p} 
& = \tilde{n} \, \tilde{T} + \frac{\sum_{i=1}^{N} \sum_{j=i+1}^{N} \mathbf{\tilde{r}}_{ij} \cdot \mathbf{\tilde{f}}_{ij}}{3\tilde{L}^3} ~.
}
Here $\mathbf{\tilde{r}}_{ij} = \mathbf{\tilde{r}}_i - \mathbf{\tilde{r}}_j$ and $\mathbf{\tilde{f}}_{ij} = -\mathbf{\tilde{f}}_{ji}$ is the force exerted by particle $j$ on particle $i$.
 The system pressure is thus evaluated as time average of Eq.~\eqref{eq:pirtheo}.

\begin{figure}[t]

    \includegraphics[width=.45\textwidth]{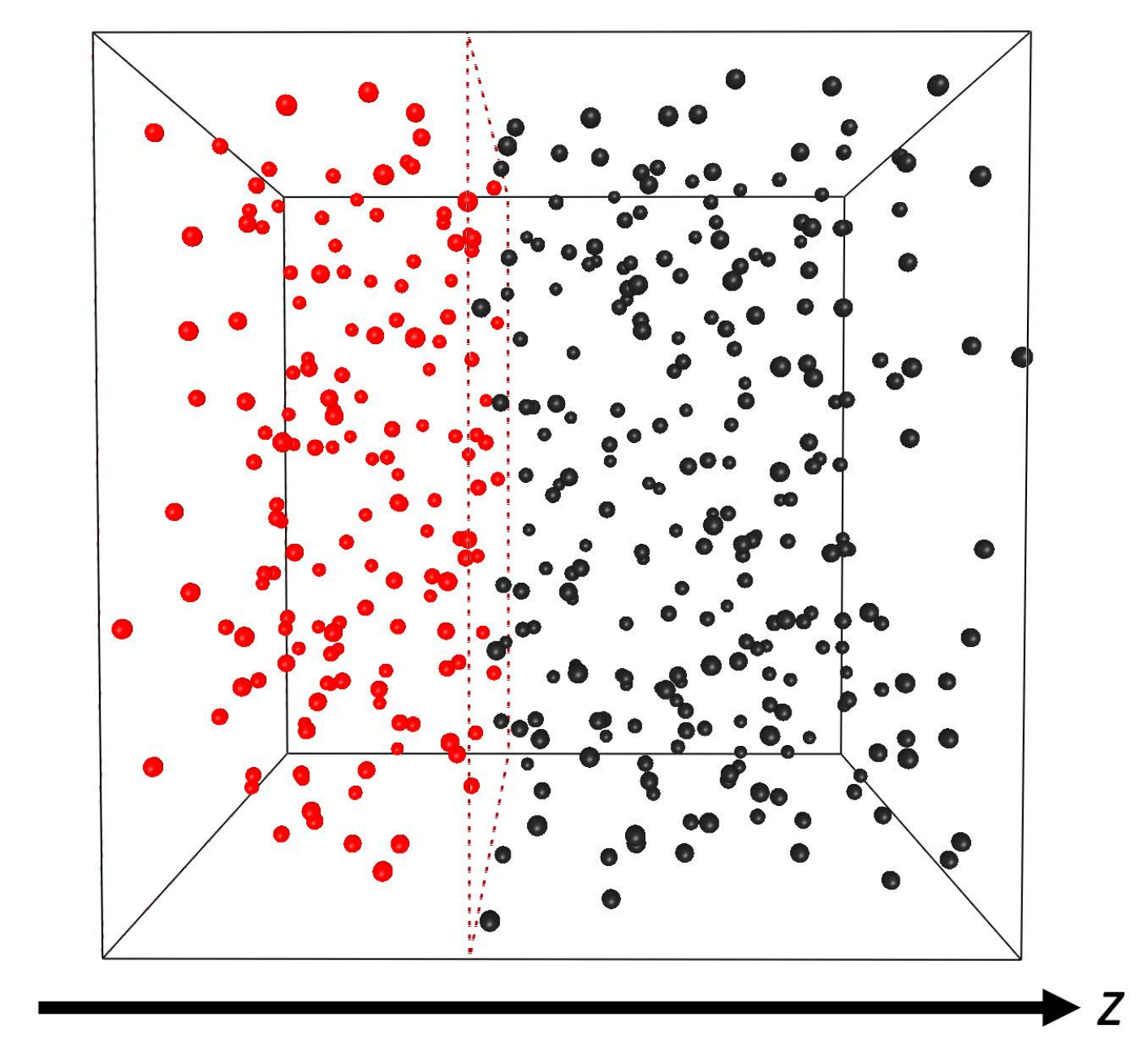} 
    \caption{
    Snapshot of a molecular dynamics simulation at $\tilde T = 1.4$ and $\tilde n = 0.3$ of the system of $N = 400$
    LJ particles, depicting the simulation box and the subvolume $\tilde z<0.4 \tilde L$  along the longitudinal coordinate.
    The red~(gray) and black spheres correspond to the LJ
    particles inside and outside the subvolume, respectively.
    }
    \label{fig:snapshot}
\end{figure}

Equation~\eqref{eq:ergod} is also used to calculate the variance of particle number fluctuations~\eqref{eq:w} by calculating $\mean{N}$ and $\mean{N^2}$ as time averages.
We study particle number fluctuations in various subsystems of the total system, namely in the coordinate space by either performing cuts $x < x^{\rm cut}$, $y < y^{\rm cut}$, or $z < z^{\rm cut}$ on the particle coordinates.
See Fig.~\ref{fig:snapshot} for an illustration of the subvolume along the longitudinal coordinate.
In addition, we also study fluctuations in the momentum space by performing cuts $|v_z| < v_z^{\rm cut}$ on the longitudinal velocity of particles.

The MD simulations are performed utilizing CUDA-enabled GPUs, which allows one to significantly speed-up the simulations relative to CPU.
The code we use is open source and available via~\cite{LJgithub}.

\section{Results}
\label{sec:results}

\begin{figure}[t]
    \includegraphics[width=.49\textwidth]{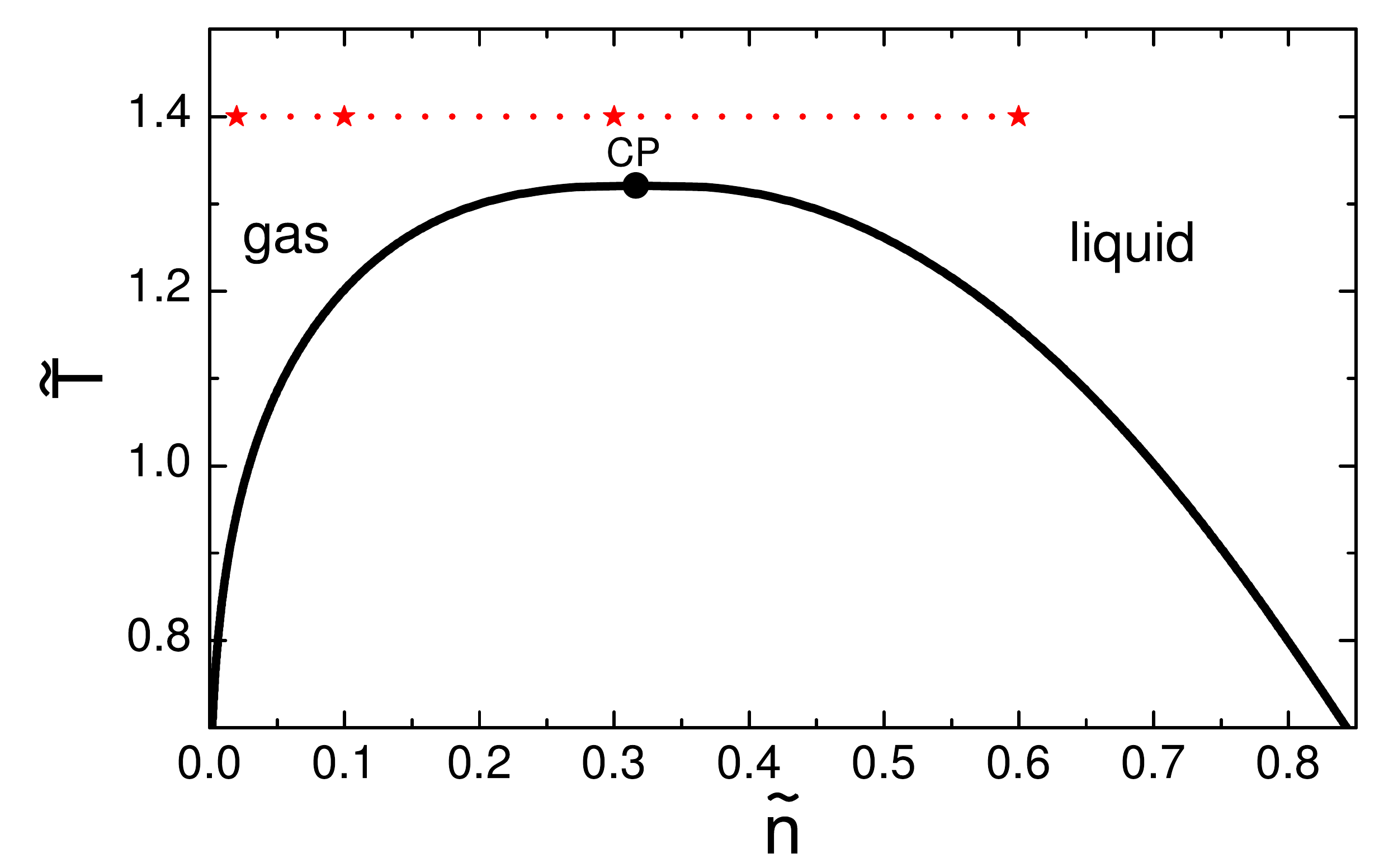} 
    \caption{
    The phase diagram of the LJ
    fluid in $\tilde T$-$\tilde n$ coordinates showing the coexistence region of the liquid-gas phase transition and the associated CP.
    The coexistence line is taken from Ref.~\cite{doi:10.1021/acs.jcim.9b00620}.
    The dotted line corresponds to the range of the $\tilde n$ values along the $\tilde T = 1.4$ isotherm used for calculations of the equation of state properties within the canonical-like ensemble~(Sec.~\ref{sec:MDEoS}).
    The red stars correspond to the $\tilde n$ values where calculations of particle number fluctuations were performed within the microcanonical ensemble~(Sec.~\ref{sec:MDflukes}).
    Note that the LJ
    fluid also exhibits phase transitions to various solid phases at $\tilde T < 0.7$ and/or $\tilde{n} > 0.85$ that are not shown in this figure.
    }
    \label{fig:LJdiagram}
\end{figure}

We focus the present study on a single isotherm corresponding to a temperature value $\tilde{T} = 1.4$. This choice is motivated by the following considerations.
On one hand, $\tilde{T} = 1.4$ is only slightly above the critical temperature of $\tilde{T}_{\rm c} \approx 1.321$, thus the effects of the CP
on particle number fluctuations should be evident along this isotherm~(see Fig.~\ref{fig:LJdiagram}).
On the other hand, given that $\tilde{T} = 1.4$ corresponds to a supercritical temperature, the system is expected to be uniform, with no mixed phase formation occurring.

\subsection{Equation of state}
\label{sec:MDEoS}

As the first step, we determine the equation of state along the $\tilde{T} = 1.4$ isotherm, namely the dependence of the pressure $\tilde{p}$ and energy per particle $\tilde{U} / N$ on the density $\tilde{n}$.
In order to do that, we run the simulations in the canonical-like ensemble, which preserves the input value of the temperature $\tilde{T}$ throughout the simulations, in particular during the equilibration stage.\footnote{In contrast, the microcanonical ensemble conserves the energy rather than the temperature. This makes it challenging to prepare the initial state with the correct temperature in the microcanonical simulation because the kinetic temperature can change considerably during the equilibration stage.}
Both the pressure and the energy are calculated as time averages~[Eq.~\eqref{eq:ergod}], with the virial theorem expression~[Eq.~\eqref{eq:pirtheo}] used for the former.

\begin{figure*}[t]
    \includegraphics[scale=0.49]{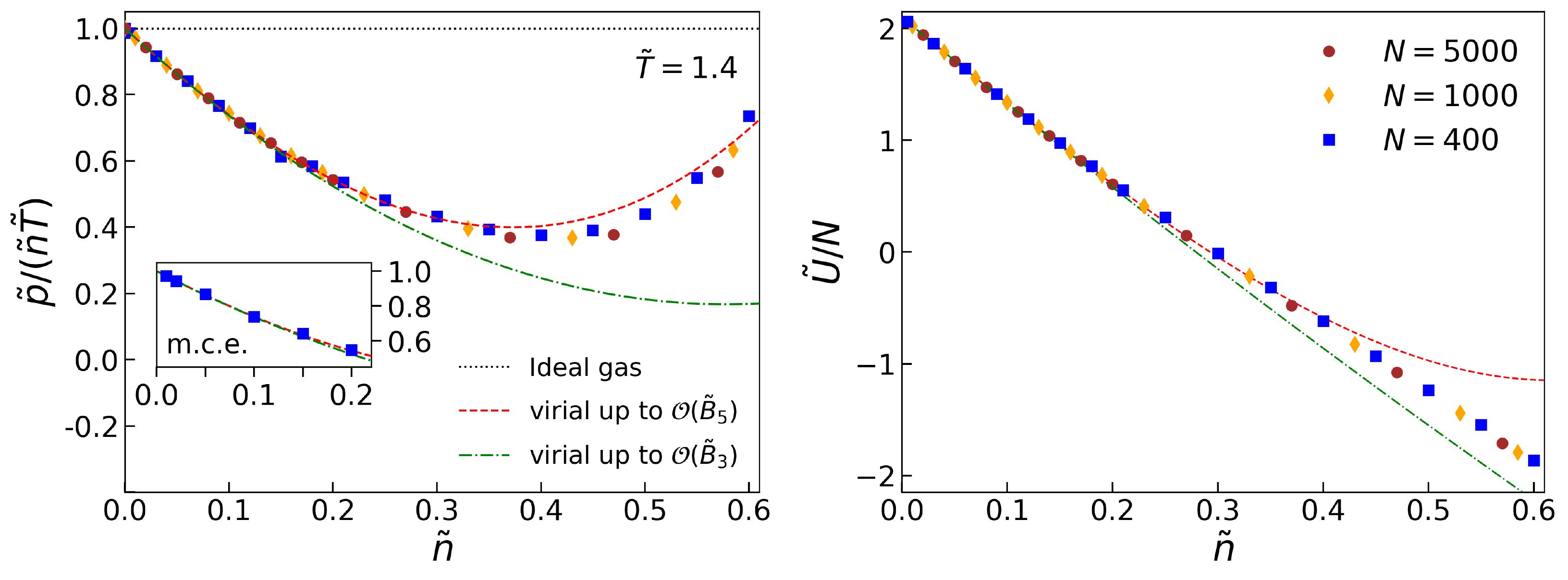} 
    \caption{
    Density dependence of the compressibility factor $Z = \tilde{p} / (\tilde{n} \tilde T)$~(left panel) and the energy per particle $\tilde{U} / N$~(right panel) along the isotherm $\tilde{T} = 1.4$ as calculated through canonical-like ensemble MD
    simulations for $N = 400$, $1000$, and $5000$ particles.
    The dash-dotted green and dashed red lines correspond to the expectations based on the virial expansion truncated at $\mathcal{O}(\tilde B_3)$ and  $\mathcal{O}(\tilde B_5)$, respectively.
    The inset in the left panel corresponds to calculations within the micro-canonical ensemble~(m.c.e.).
    }
    \label{fig:P}
\end{figure*}

Figure~\ref{fig:P} depicts the resulting density dependence of the compressibility factor~$Z$ and the energy per particle $\tilde{U} / N$.
The calculations were performed for different values of the total number of particles, $N = 400$, $1000$, and $5000$, and these different cases are depicted by the different symbols.
The simulation results are compared to the virial expansion of the LJ equation of state along the same isotherm, which is truncated at $\tilde{B}_3$~(dashed green line) or $\tilde{B}_5$~(dashed red line).
The MD simulations agree with the $\mathcal{O}(\tilde{B}_3)$ virial expansion at low densities~($\tilde{n} \lesssim 0.2$) for all used values of $N$, validating the accuracy of simulations. The agreement at higher densities is improved if more terms are incorporated into the virial expansion, as evidenced by the $\mathcal{O}(\tilde{B}_5)$ calculation for $Z$~(dashed red line).

As a further cross-check of the accuracy of MD simulations in the canonical-like ensemble, we also ran simulations in the microcanonical ensemble using the computed values of $\tilde{U} / N$ as input into the initial conditions. 
The microcanonical simulations yielded the same results for the pressure, while the average kinetic temperature in these simulations is consistent with $\tilde{T} = 1.4$.
We also checked that the results for $Z$ are reproduced by a different method, through the integration of the radial distribution function, as discussed in Appendix~\ref{app:RDF}.

\subsection{Fluctuations}
\label{sec:MDflukes}

\subsubsection{Grand-canonical limit}

We focus on the scaled variance of particle number fluctuations~[Eq.~\eqref{eq:w}].
Before calculating the fluctuations directly, we first analyze the analytical expectations for the behavior of $\omega$ along the isotherm as function of density.
In the low-density limit, $\tilde{n} \to 0$, the system is expected to approach the ideal gas limit, where $\omega \to 1$.
At large densities the system behavior may be dominated by the short-range repulsion, which would be expected to suppress the fluctuations~\cite{Vovchenko:2015xja}.
At intermediate densities, $\tilde{n} \sim \tilde{n}_c \sim 0.316$, the system is in the vicinity of the CP, where the variance is expected to grow large.
Thus, at $\tilde{T} = 1.4$ 
and $\tilde{n} \sim \tilde{n}_c$ one would expect a peak in the density dependence of $\omega$.

\begin{figure}[t]
    \includegraphics[width=.49\textwidth]{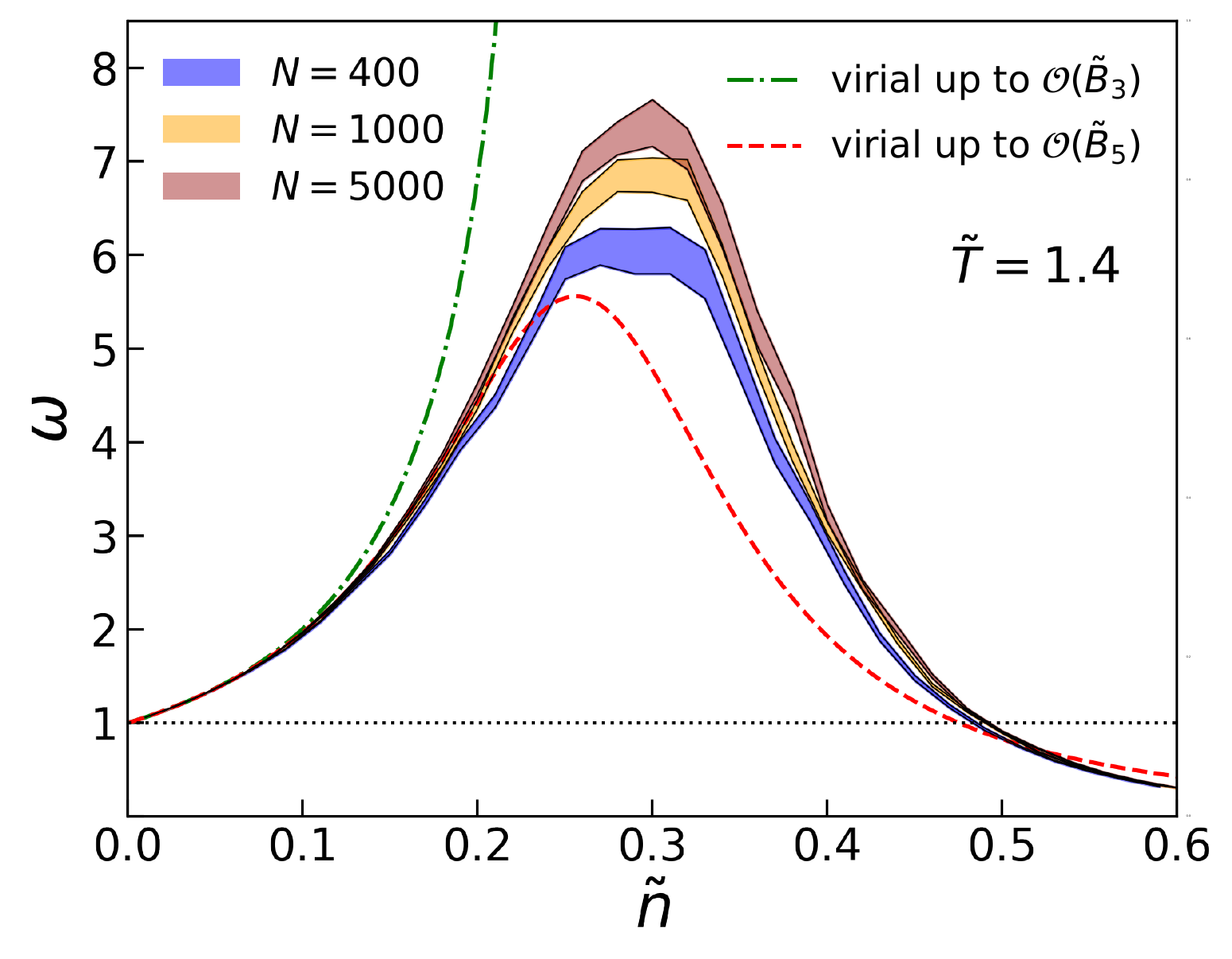} 
    \caption{
    The scaled variance of particle number fluctuations in grand-canonical ensemble along the $\tilde T = 1.4$ isotherm calculated via Eq.~\eqref{eq:w} using molecular dynamics data for $Z$.
    Different bands corresponds to different number of particles used in simulations: blue~(lower) for $N = 400$, yellow~(middle) for $N = 1000$, and red~(upper) for $N = 5000$.
    The dash-dotted green and dashed red lines correspond to the expectations based on the virial expansion truncated at $\mathcal{O}(\tilde B_3)$ and  $\mathcal{O}(\tilde B_5)$, respectively.
    }
    \label{fig:wi}
\end{figure}

These qualitative expectations can be tested with our MD results via Eq.~\eqref{eq:w}, through the use of the density dependence of the compressibility factor $Z$ computed with MD.
The derivative $(\partial \tilde{Z} / \partial \tilde{n})_{\tilde{T}}$ is calculated through the finite difference method, using the first-order central difference.\footnote{The uncertainties in $\omega$ are calculated through error propagation of $\tilde{Z}$ in the finite difference expressions.}
The results of this procedure, applied to simulations with different $N$ values, are depicted in Fig.~\ref{fig:wi}.
The calculations show that $\omega$ is peaked at $\tilde{n} \approx 0.3$, approaches unity -- the Poisson limit -- for $\tilde{n} \to 0$, while at high densities it is suppressed and approaches zero.
The results do slightly depend on $N$~(or, equivalently, the system volume $\tilde V = N / \tilde n$), especially in the vicinity of the peak, where the $N = 400$ calculations show a smaller peak compared to $N = 1000$ and $N = 5000$.
This reflects the system-size dependence in the evaluation of $d \tilde{p} / d \tilde{n}$, which is most pronounced near $\tilde{n} \sim 0.3$ where $d \tilde{p} / d \tilde{n}$ attains small values reflecting large correlation length.
The MD
results are compared to the virial expansion~[Eq.~\eqref{eq:wvir}] truncated at $\mathcal{O}(\tilde{B}_5)$, which is depicted in Fig.~\ref{fig:wi} by the dashed red line.
The MD calculations agree with the virial expansion at $\tilde{n} \lesssim 0.1$ for all the values of $N$ considered, while for the simulations with $N = 1000$ and $N = 5000$ the agreement range is larger, up to $\tilde{n} \lesssim 0.2$.

The results for $\omega$ are in qualitative agreement with the analytic predictions of the van der Waals model as detailed in Appendix~\ref{app:vdWcompare}.

\subsubsection{Coordinate space subsystem}

Next, we look at the fluctuations of particle number that occur throughout the MD simulation. 
The total particle number in the entire volume is fixed. We thus study the behavior of particle number in a subsystem of the whole system where it can fluctuate.
Then, we analyze whether the results can be connected to the grand-canonical scaled variance shown in Fig.~\ref{fig:wi}.

In contrast to the mean quantities, the behavior of fluctuations depends on the choice of simulation ensemble. For this reason, the calculations of particle number fluctuations that occur throughout the MD simulations are performed in the microcanonical ensemble rather than in the canonical-like ensemble that we used before.
As discussed above, we use the values of the energy per particle $\tilde{U} / N$ computed previously in the canonical-like ensemble as input into the microcanonical ensemble simulation in order for our simulations to correspond to the desired temperature of $\tilde T = 1.4$ at given particle number density.

First, we analyze the fluctuations in coordinate space subsystems.
In Ref.~\cite{Vovchenko:2020tsr} it was shown that these fluctuations can be related to the grand-canonical susceptibilities in the large volume limit. Namely, the scaled variance reads
\eq{
\omega^{\rm coord} = (1-\alpha) \, \omega^{\rm gce}~.
}
Here $\omega^{\rm gce}$ is the grand-canonical scaled variance~[Eq.~\eqref{eq:w}] and $\alpha$ is the fraction of the total volume occupied by the subvolume.

Here we define the subvolume by performing cuts $w < w^{\rm cut}$, where $w$ is either $\tilde x$, $\tilde y$, or $\tilde z$\footnote{Note that in our notation the  coordinate values vary in the range $0 < \tilde x,\tilde y, \tilde z < \tilde{L}$.}.
It follows that $\alpha = w^{\rm cut} / \tilde{L}$.
Due to the cubic symmetry of our simulation setup, the results are expected to be identical for the same value of $\alpha$ regardless of which coordinate is chosen, as long as $\tilde \tau$ is sufficiently large to ensure the ergodicity.
We verified explicitly that, for the same value of $\alpha$, the results for $\omega^{\rm coord}$ using either of the three Cartesian coordinates are consistent with each other within the statistical uncertainty.
Thus, in order to reduce the total statistical error, we averaged the results over the calculations utilizing the cuts in $\tilde x$, $\tilde y$, and $\tilde z$.
Furthermore, the variance $\mean{\Delta N^2}$ is symmetric with respect to a change $\alpha \to 1 - \alpha$~\cite{Bzdak:2017ltv}. We thus symmetrize our results with respect to $\alpha \to 1 - \alpha$ to further decrease the statistical error.

\begin{figure*}[t]
\includegraphics[scale=1.15]{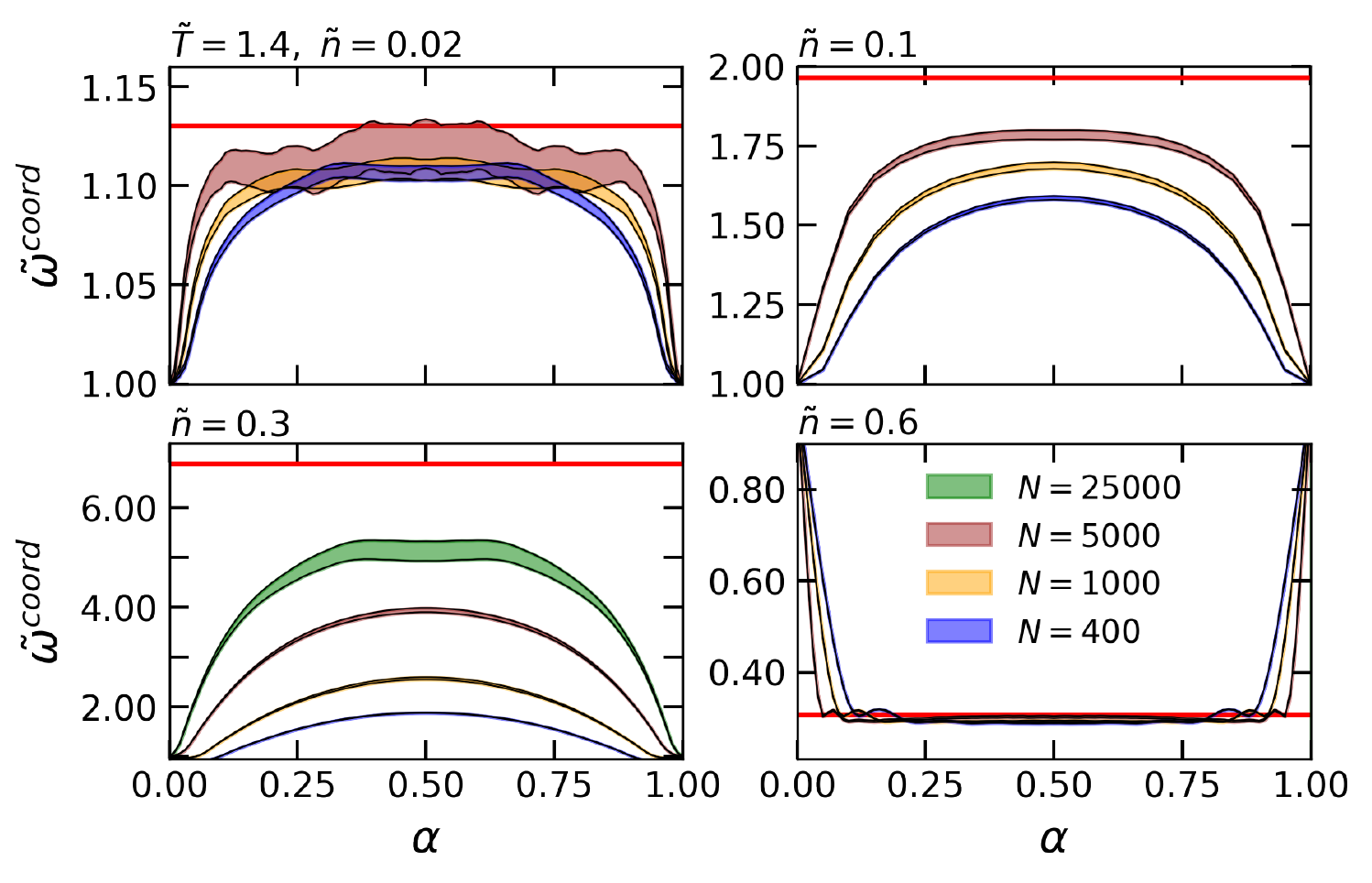}  
\caption{
Scaled variance of particle number fluctuations $\tilde \omega^{\rm coord}$ inside coordinate space subvolumes calculated through MD simulations in the micro-canonical ensemble for different values of the density $\tilde{n}$ and number of particles $N$.
The results for $\tilde{\omega}^{\rm coord}$ corrected for global particle number conservation through the $(1-\alpha)$ factor   are presented as a function of subvolume fraction $\alpha$.
The widths of the bands correspond to the statistical uncertainties and their colors to the values of $N$, which vary as $N = 400$, $N = 1000$, $N = 5000$ and, in the case of $\tilde n = 0.3$, also $N = 25000$.
The larger $N$ is, the closer the corresponding bands are to the expected thermodynamic limit, depicted by the horizontal red lines.
}
\label{fig:fsw0}
\end{figure*}

Figure~\ref{fig:fsw0} depicts the MD results for the scaled variance $\tilde{\omega}^{\rm coord} \equiv \omega^{\rm coord} / (1-\alpha)$ corrected for particle number conservation as function of $\alpha$ for different values of the density $\tilde{n}$ along the isotherm $\tilde{T} = 1.4$.
For $\tilde{n} = 0.02$, the system is dilute and exhibits properties similar to an ideal gas of particles at the same temperature and density.
In the grand-canonical limit, the scaled variance of particle number fluctuations is expected to show a slight enhancement over the Poisson limit, namely $\omega^{\rm gce} \simeq 1.126$, as follows from both the virial expansion and the MD based calculations of $\omega^{\rm gce}$ via Eq.~\eqref{eq:w} shown in Fig.~\ref{fig:wi}.
The MD simulation results for $\tilde{\omega}^{\rm coord}$ lie in the range between 1 and $\omega^{\rm gce} \simeq 1.126$, i.e. they do not exceed the grand-canonical limit.
These results approach the grand-canonical limit if $\alpha$ is not too close to 0 or 1, as well as when the number of particles $N$ is increased.

For $\tilde n = 0.1$ the effects of interactions are more prominent, with the grand-canonical scaled variance $\omega^{\rm gce} \simeq 1.97$ being almost double the Poisson value.
The finite-size effects are also more prominent here, namely, they suppress the fluctuations, which is shown by the MD simulations results being consistently below $\omega^{\rm gce}$, even for $N = 5000$. 
We do observe, however, that $\tilde{\omega}^{\rm coord}$ is larger for larger $N$ and the trend is consistent with approaching the grand-canonical limit as $N \to \infty$.

At $\tilde{n} = 0.3$~(and $\tilde{T} = 1.4$) the system is located close to the CP at $\tilde{n}_c \approx 0.316$ and $\tilde{T}_c \approx 1.321$.
This is characterized by large grand-canonical fluctuations of particle number, namely $\omega^{\rm gce} \simeq 7-7.5$~(see Fig.~\ref{fig:wi}).
As seen from Fig.~\ref{fig:wz}, large fluctuations of $\tilde{\omega}^{\rm coord}$ are also observed in MD simulations, with the maximum values reached at $\alpha = 0.5$.
The results exhibit strong system-size dependence, with the magnitude of $\tilde{\omega}^{\rm coord}$ depending strongly on the total number of particles in the system.
For instance, at $N = 5000$ the maximum value of $\tilde{\omega}^{\rm coord}$ is still about half that of the expected thermodynamic limit and even $N = 25000$ is not sufficient to reach the limit. \footnote{Note that even though we employ periodic boundary conditions, this does not lead to a possible double counting of the CP effects from multiple boxes. This is due to the minimum-image convention scheme that we use, where each particle interacts with only a single (the closest one) image of every other particle.}
Nevertheless, the results clearly show that the CP does lead to sizeable fluctuations of particle number in finite systems, justifying the search for large fluctuations as a signature of criticality.

The large density case, $\tilde{n} = 0.6$, is qualitatively different from the other cases. Here the fluctuations are suppressed relative to the Poisson baseline, with the grand-canonical scaled variance being equal to $\omega^{\rm gce} \simeq 0.30-0.31$.
This suppression is also observed in MD, with the simulation results saturating at $\tilde{\omega}^{\rm coord} \approx 0.3$ in a broad interval around $\alpha = 0.5$. The results exhibit only mild system-size dependence and the obtained values are consistent with the grand-canonical expectation in the thermodynamic limit.
We also observed that the ergodicity is reached considerably faster, with $\tilde{\tau} = 10000$ being sufficient to obtain accurate results for $\tilde{\omega}^{\rm coord}$, which is about an order of magnitude lower value compared to that required at lower densities.

In all cases the scaled variance tends to unity in the limit $\alpha \to 0$. This is the expected result reflecting the so-called ``Poissonization'' of fluctuations in small volumes~(acceptance)~\cite{Bzdak:2012ab}, when the system size becomes smaller than the correlation length.
In the opposite limit, $\alpha \to 1$, the scaled variance vanishes due to the global conservation of particle number, $\omega^{\rm coord} \to 0$. The scaled variance corrected for global conservation, $\tilde{\omega}^{\rm coord}$, exhibits the same behavior as in the $\alpha \to 0$ limit due to the symmetry between the subsystem and the complement~\cite{Bzdak:2017ltv}.

\subsubsection{Momentum space subsystem}

Here we study the behavior of fluctuations in the momentum space, by performing a cut $|v_z| < v_z^{\rm cut}$ on the longitudinal velocity of particles.
Such a procedure resembles fluctuation measurements in heavy-ion collision experiments, where only the momenta, not the coordinates, of particles can be determined.
Interactions between particles in the LJ fluid depend only on their coordinates, but not the momenta.
In fact, it can be shown that the multiparticle momentum distribution function in the canonical ensemble factorizes into a product of single-particle Maxwell-Boltzmann distribution functions, whereas all the effects of interactions are washed out by integrating over the coordinates of all the particles.
Therefore, the scaled variance of particle number fluctuations in the momentum space is expected, in the canonical ensemble, to reduce to the binomial distribution stemming from global particle number fluctuations~\cite{Bzdak:2012an,Savchuk:2019xfg}, $\omega^{\rm mom,ce} = 1 - \alpha$, where $\alpha = \mean{N_{\rm acc}} / N$ and $\mean{N_{\rm acc}}$ is the mean number of particles in the momentum acceptance.

In the microcanonical ensemble, however, the fluctuations can additionally be affected by exact conservation of energy-momentum.
One can derive the following baseline for $\tilde{\omega}^{\rm mom,mce}_{\rm id} = \omega^{\rm mom,mce}_{\rm id} / (1-\alpha)$ in the framework of ideal gas of particles in the microcanonical ensemble in thermodynamic limit, assuming, as before, that the momentum subspace corresponds to a cut $|v_z| < v_z^{\rm cut}$
\eq{\label{eq:wmom}
\tilde{\omega}^{\rm mom,mce}_{\rm id} = 1 - \frac{2 [\operatorname{erf}^{-1} (\alpha)]^2 e^{-2[\operatorname{erf}^{-1} (\alpha)]^2}}{3\pi\alpha (1 - \alpha)}~.
}
The details of the derivation are given in Appendix~\ref{app:microcan}, where it is also shown that Eq.~\eqref{eq:wmom} is quantitatively accurate for systems of 400 or more particles.
In the case of an interacting system, like the LJ fluid, one can expect corrections to Eq.~\eqref{eq:wmom} due to the influence of the interaction energy on the total energy-momentum conservation.

\begin{figure}[t]
    \includegraphics[width=.49\textwidth]{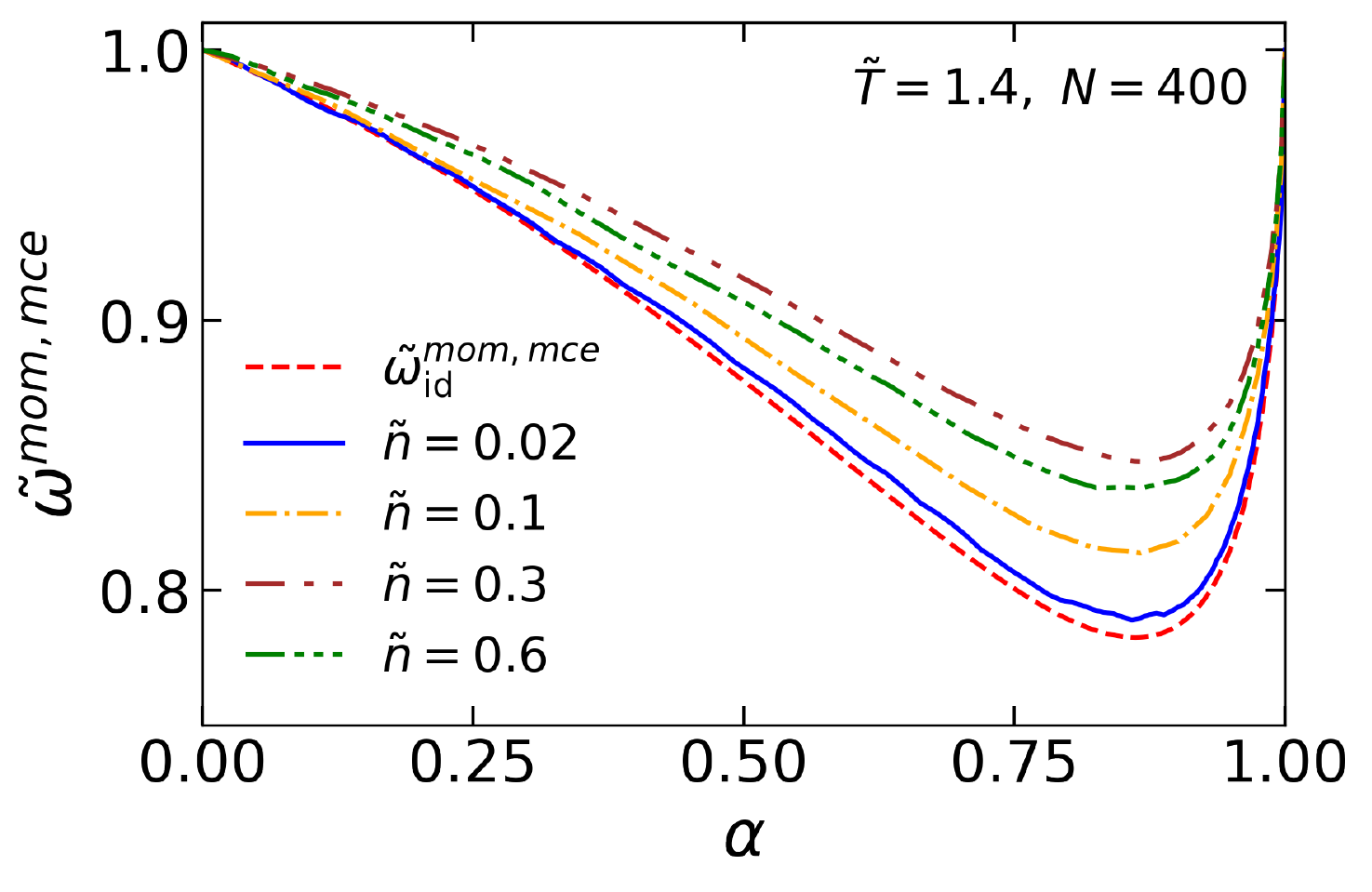}   
    \caption{
    Scaled variance of particle number fluctuations $\tilde \omega^{\rm mom,mce}$ in the momentum space subsystem defined by a cut $|v_z| < v_z^{\rm cut}$ in the longitudinal velocity, as obtained from molecular dynamics simulations in the microcanonical ensemble for $N = 400$ particles at different densities.
    The results are presented as a function of subsystem fraction $\alpha \equiv \mean{N_{\rm acc}} / N$ and corrected for global particle number conservation through the $(1-\alpha)$ factor.
    The dashed red line corresponds to the ideal gas limit given by Eq.~\eqref{eq:wmom}.
    }
    \label{fig:wz}
\end{figure}	

Figure~\ref{fig:wz} shows the results of MD simulations for $\tilde{\omega}^{\rm mom,mce}$ as a function of $\alpha$ for different values of particle number density.\footnote{The value of $\alpha = \mean{N_{\rm acc}} / N$ is regulated by the choice of $v_z^{\rm cut}$.}
The results are compared with the expected low-density~(ideal gas) limit given by Eq.~\eqref{eq:wmom},  shown by the dashed red line in Fig.~\ref{fig:wz}.
The MD calculations at the lowest considered density~($\tilde{n} = 0.02$) are close to the low-density limit, while the calculations at larger densities show slightly larger deviations, but the same qualitative behavior: a nonmonotonic dependence of $\tilde{\omega}^{\rm mom,mce}$ on $\alpha_N$ with a minimum at $\alpha_N \approx 0.85$.
We also observe that $\tilde{\omega}^{\rm mom,mce}$ never exceeds unity, even in the vicinity of the critical density, $\tilde{n} = 0.3$, thus, the CP signal in particle number fluctuations is essentially washed out when one analyzes them in the momentum space.
The reason is that coordinates and momenta of particles are uncorrelated in our box simulation, thus, the enhancement of particle number fluctuations predicted by the theory for coordinate space subvolumes does not translate into the momentum space.

The results have relevance for the QCD CP search in heavy-ion collisions via the analysis of event-by-event fluctuations. 
Due to experimental limitations, it is only possible to measure the momenta of hadrons created in heavy-ion collisions, but not their coordinates at freeze-out, thus, the analysis is necessarily performed in the momentum space.
Our results point to the challenges associated with the analysis of fluctuations in the momentum space: in the absence of correlations between the momenta and coordinates of particle it is extremely challenging to observe fluctuation signals of the CP in particle number distributions.
It should be noted, however, that the system created in heavy-ion collisions differs from the one studied here. 
Instead of a system of classical particles in a box with periodic boundary conditions, heavy-ion collisions create a droplet of QCD fluid that expands into vacuum and hadronizes.
This leads to the development of collective flow velocities that generate correlations between the coordinates and momenta of hadrons at the freeze-out stage.
In the limiting case of Bjorken flow, the correlation is one-to-one between the longitudinal coordinates and collective velocities of particles, with their final velocities affected additionally only by thermal smearing~\cite{Ling:2015yau,Vovchenko:2020kwg}.

In this regard, MD simulations can be extended to make them more appropriate for heavy-ion applications.
This can be achieved by letting the thermalized system expand, possibly with boosted velocities to account for the effect of collective flow, and then analyzing the fluctuations in momentum space as ensemble averages.
Furthermore, it may be important to incorporate explicitly the formation of composite bound states like light nuclei, which are formed in abundance in heavy-ion collisions at intermediate energies. 
Furthermore, by looking at ensemble averages rather than time averages, one can also study the dynamics of the equilibration stage for event-by-event fluctuations.
These extensions will be the subject of future studies.

\section{Summary and outlook}
\label{sec:summary}

In this work we studied particle number fluctuations in and out of the vicinity of a critical point microscopically, by utilizing molecular dynamics simulations of the Lennard-Jones fluid. 
The simulations were performed in a box with periodic boundary conditions, naturally incorporating effects like physics of the correlation length, exact conservation laws, and finite size.

To study the effect of the proximity of the CP, we performed calculations along an isotherm $\tilde{T} = 1.4 \simeq 1.05 \tilde{T}_c$, i.e. slightly above the critical one, for different values of particle number density. 
The simulations were performed in two steps.
First, simulations in a canonical-like ensemble were performed to map the temperature $\tilde{T} = 1.4$ to the corresponding mean energy per particle, $\tilde{U} / N$, and pressure, $\tilde{p}$, for each considered value of the particle number density.
It has been checked that the obtained results are consistent at low densities with the analytic expectations based on the virial expansion.

Then, microcanonical ensemble simulations along the same isotherm were performed, using the computed $\tilde{U} / N$ values as input. We studied in detail the behavior of the scaled variance $\omega^{\rm coord}$ of particle number fluctuations in various coordinate space subsystems. 
The fluctuations have been computed through Eq.~\eqref{eq:w} via time averages of $\mean{N^2}$ and $\mean{N}$, while the coordinate space subsystems were defined via cuts in one of the Cartesian coordinates, i.e. $x < x^{\rm cut}$, $y < y^{\rm cut}$, or $z < z^{\rm cut}$.
It has been checked that the results are consistent within errors for all three choices of the Cartesian coordinate, thus, to minimize the values of the statistical error, the results were averaged over the three choices.
The scaled variance corrected for global conservation, $\tilde{\omega}^{\rm coord} \equiv \omega^{\rm coord} / (1-\alpha)$, is expected to coincide with the grand-canonical scaled variance in the thermodynamic limit, as shown earlier in Refs.~\cite{Vovchenko:2020tsr,Vovchenko:2020gne}.
The MD simulation results approach the thermodynamic limit as the system volume $V$~(or, equivalently, the total number of particles $N$) increases, as shown in Fig.~\ref{fig:wi}.
We do observe that the simulations generally yield smaller values of $\tilde{\omega}^{\rm coord}$ compared to the thermodynamic limit, reflecting the system-size effect, especially in the vicinity of the critical particle number density $\tilde{n} = 0.3 \simeq 0.95 n_c$.
This observation is consistent with the earlier study performed in the framework of the van der Waals model in Ref.~\cite{Poberezhnyuk:2020ayn}.
One sees that, even though the finite-size effects are significant near the CP, the strong enhancement of fluctuations is indeed shown to be a viable signature of the CP, as long as the fluctuations are analyzed in coordinate space subvolumes.

We then analyzed the behavior of fluctuations in momentum space, by performing a cut $|v_z| < v_z^{\rm cut}$ on the longitudinal velocities of particles, which reflects better the conditions realized in heavy-ion collision experiments.
One sees that the strong enhancement of fluctuations due to the CP is not present in momentum space, and the qualitative behavior of $\tilde{\omega}^{\rm mom}$ is determined by the effect of exact energy conservation.
The reason is that the momenta and coordinates of the Lennard-Jones particles are uncorrelated in equilibrium, reflecting the fact that the canonical partition function of the system factorizes into a momentum and coordinate dependent parts.

It should be noted, however, that the system created in heavy-ion collisions differs from a box with periodic boundary conditions that was studied here. For one thing, fluctuations are analyzed in the experiment as event-by-event~(ensemble) averages, rather than time averages calculated here. And while the two are expected to coincide within errors due to the ergodic hypothesis, it can be instructive to explicitly verify that this is the case.
An even more important difference is that the system created in heavy-ion collisions is not static, but expands into vacuum and is usually characterized by the presence of sizable collective flow at the freeze-out stage. This, in turn, generates a degree of correlation between the freeze-out coordinates and momenta of hadrons.
In that regard, it would be interesting to study the expansion of an equilibrated Lennard-Jones fluid, and how this may translate the coordinate space fluctuations into the momentum space ones. 
We plan to study this question in a separate publication.

The analysis can be extended to higher-order cumulants of particle number like skewness and kurtosis, as these non-Gaussian measures are expected to be even more sensitive probes of the (QCD) CP compared to the variance. 
Such calculations, however, are likely to require considerably more computing resources, as the statistical error is typically larger for cumulants of higher order~\cite{Bzdak:2018zdg}.

Another interesting possibility is the first-order phase transition at subcritical temperatures, with the associated mixed phase formation and its possible signatures in fluctuation observables. The MD simulations of the Lennard-Jones fluid describe the mixed phase formation microscopically and thus are well-suited for such studies.
Finally, transport properties like the diffusion coefficient, as well as shear and bulk viscosity, can be calculated and their behavior near the CP elaborated.

\begin{acknowledgments}

This work received support through the U.S. Department of Energy, 
Office of Science, Office of Nuclear Physics, under contract number 
DE-AC02-05CH11231231 and within the framework of the
Beam Energy Scan Theory (BEST) Topical Collaboration.
V.V. acknowledges the support through the
Feodor Lynen Program of the Alexander von Humboldt
foundation.  M.I.G  acknowledges the  support  from the National Academy of Sciences of Ukraine, Grant No. 0121U112254.
This research used the Lawrencium computational cluster resource provided by the IT Division at the Lawrence Berkeley National Laboratory.
\end{acknowledgments}

\appendix

\begin{widetext}

\section{Virial coefficients}
\label{app:virial}

The 2nd virial coefficients of the LJ fluid can be calculated analytically~\cite{VARGAS200192}. It reads
\eq{
\tilde{B}_2(\tilde{T}) = \frac{\pi^2 \sqrt{2}e^{1/2\tilde{T}}}{3\tilde{T}}\left[I_{3/4}\left(\frac{1}{2\tilde{T}}\right)+I_{-3/4}\left(\frac{1}{2\tilde{T}}\right)-I_{1/4}\left(\frac{1}{2\tilde{T}}\right)-I_{-1/4}\left(\frac{1}{2\tilde{T}}\right)\right]~.
}
Here $I_\alpha$ is the modified Bessel function of the first kind.
The data for the temperature dependence of the virial coefficients with $3\leq i \leq 6$ can be parameterized in the following form~\cite{Gott}:
\eq{\tilde B_i(\tilde T)= \left(\frac{\tilde T}{4}\right)^{-\frac{i-1}{4}}\left[\tilde B_i^{SS} + \sum_{k=1}^{k_i}b_{i,k}\left(\exp{\frac{c_i}{\sqrt{\tilde T}}} - 1\right)^{\frac{2k-1}{4}}\right].}
Here, explicit formula and parameters of the thermal virial coefficients are taken from \cite{Gott}. This approximation is appropriate for a broad temperature range of $0.25 < \tilde T < 25$. 

\end{widetext}

\subsection*{Virial expansion for the energy}

The total energy in the canonical ensemble is
\eq{
\tilde{U} = \tilde{F} + \tilde{T} \tilde{S},
}
where $\tilde{F}$ is the free energy and $\tilde{S} = -(\partial \tilde{F} / \partial \tilde{T})$ is the entropy.
The virial expansion for the free energy can be found by integrating the equation
$$\tilde{p} = - (\partial F / \partial V)$$
using Eq.~\eqref{eq:Pvir} for the pressure and fixing the integration constant to get the ideal gas limit for $\tilde{B}_k \to 0$:
\eq{
\tilde{F} = \tilde{F}^{\rm id} + \tilde{V} \tilde{T} \sum_{k=2}^\infty \frac{\tilde{B}_k(\tilde{T})}{k-1} \tilde{n}^{k}~.
}
Calculating the entropy $\tilde{S} = -(\partial \tilde{F} / \partial \tilde{T})$ and plugging it into the expression for the energy $\tilde{U}$ one obtains
\eq{
\tilde{U} & = \tilde{U}^{\rm id} - \tilde{V} \sum_{k=2}^{\infty} \frac{\tilde{T}^2 \tilde{B}'_k(\tilde{T})}{k-1} \tilde{n}^k \nonumber \\
& = \frac{3}{2} N \tilde{T} - \tilde{V} \sum_{k=2}^{\infty} \frac{\tilde{T}^2 \tilde{B}'_k(\tilde{T})}{k-1} \tilde{n}^k
}

\section{Radial distribution function}
\label{app:RDF}

The radial distribution function $g(\tilde r)$ describes how the (time-averaged) density of particles varies around a reference particle at $\tilde r = 0$ relative to the expectation based on the mean particle number density $\tilde n = N / \tilde{V}$.
Namely, $g(\tilde r)$ is defined such that the local particle number density at a distance $\tilde r$ from the reference particle equals $\tilde n \, g(\tilde r)$.
In the ideal gas limit, i.e. in the absence of interactions between particles, one has $g(\tilde r) = 1$ for all $\tilde r > 0$.

The presence of interactions leads to deviations of $g(\tilde r)$ from unity.
In the dilute limit one can assume that the reference particle interacts with at most one other particle. In this case $g^{\rm dil}(\tilde r)$ is given by the Bolztmann distribution involving the pair interaction energy, thus
\eq{\label{eq:gdil}
g^{\rm dil}_{\rm LJ}(\tilde r) = \exp \left[-\frac{\tilde V_{\rm LJ}(\tilde r)}{\tilde T} \right]
}
is the expected low-density limit for the LJ fluid.
At larger densities the structure of $g(\tilde r)$ becomes more complicated, but still can be studied with MD simulations.

\begin{figure}[t]
    \includegraphics[width=.49\textwidth]{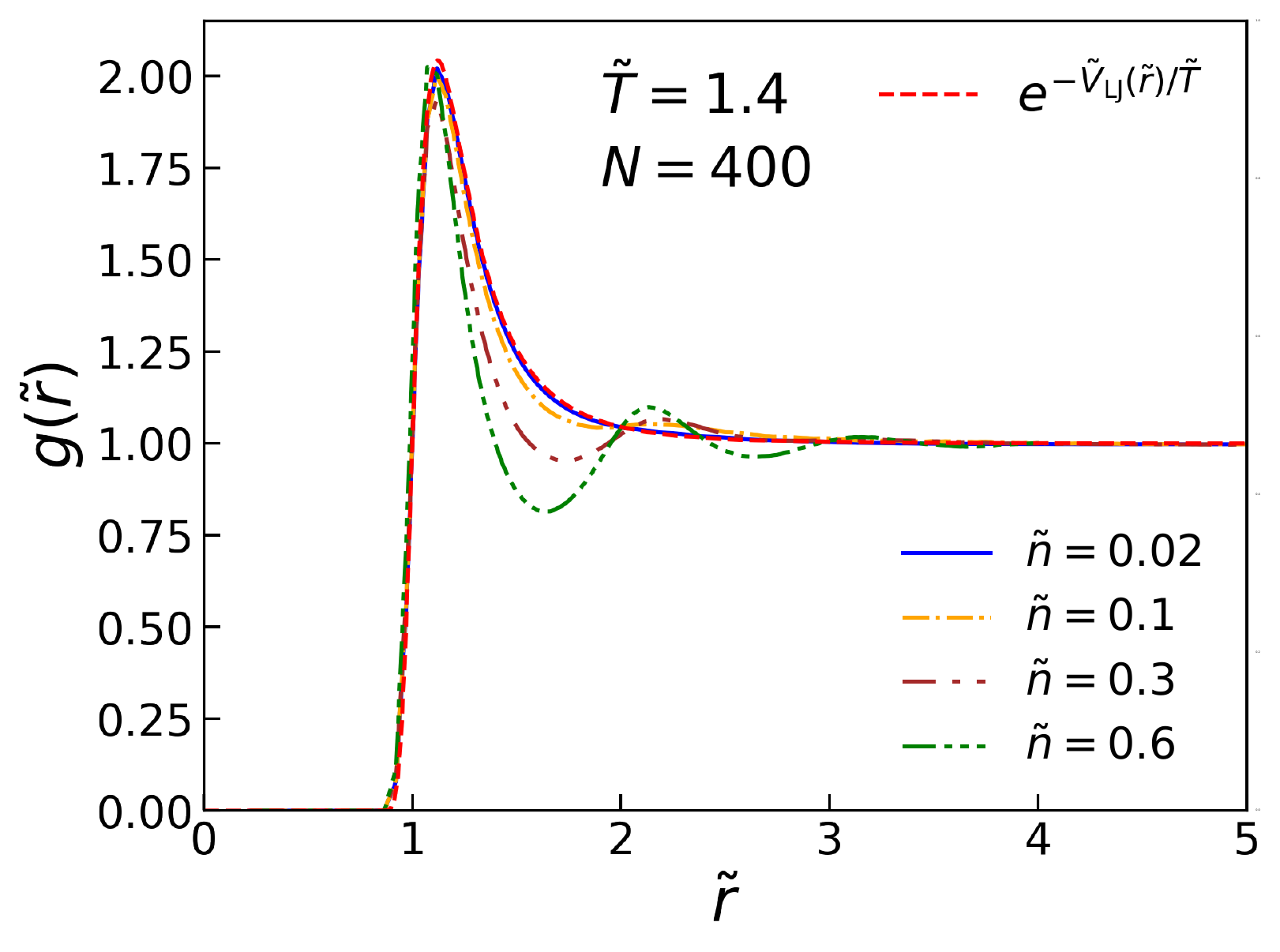}   
    \caption{
    Radial distribution function $g(\tilde r)$ of the Lennard-Jones fluid calculated with molecular dynamics simulations at $\tilde T = 1.4$ and $\tilde n = 0.02$, $0.1$, $0.3$, and $0.6$, shown by the lines of different color and style.
    The dilute limit given by Eq.~\eqref{eq:gdil} is depicted by the dashed red line.
    }
    \label{fig:RDF}
\end{figure}

Here we calculated $g(\tilde r)$ for the LJ fluid numerically, by utilizing MD simulations 
in the microcanonical ensemble at $\tilde n = 0.02$, $0.1$, $0.3$, and $0.6$ and $\tilde T = 1.4$.
This has been achieved through a (time-averaged) histogram binning of all pair distances throughout the MD simulation.
The results are depicted in Fig.~\ref{fig:RDF}.
They were obtained utilizing $N = 400$ simulations, we also checked that $N = 1000$ and $N = 5000$ simulations give essentially the same results, thus only the $N = 400$ case is shown.
For all the densities, the radial distribution function quickly drops to zero at small distances, $\tilde r \lesssim 1$, reflecting the approximately hard-core nature of the short-range repulsion given by the LJ potential that does not allow any two particles to overlap.
At large distances, $\tilde{r} \gtrsim  3$, $g(\tilde r)$ approaches unity, indicating that the influence of the reference particle on the local density of particles diminishes as the distance is increased, as expected.
The behavior of $g (\tilde r)$ at intermediate distances, $1 \lesssim \tilde r \lesssim 3$, is interesting and exhibits a notable density dependence.
At the lowest considered density, $\tilde n = 0.02$, $g (\tilde r)$ is close to the low-density limit given by Eq.~\eqref{eq:gdil} for all $\tilde r$.
As the density is increased, deviations from Eq.~\eqref{eq:gdil} become more evident.
In particular, for $\tilde n = 0.6$, $g (\tilde r)$ exhibits multiple peaks and dips, indicating the formation of long range order at high densities.

The radial distribution function can be used to evaluate the equation of state.
In particular, the pressure of a system interacting through a central pair potential, like the LJ potential $\tilde V_{\rm LJ}(\tilde r)$, reads~\cite{frenkel2001understanding}
\eq{\label{eq:Prdf}
\tilde{p} = \tilde n \tilde T - \frac{2}{3} \pi \tilde n^2 \, \int_0^\infty dr \, r^3 \, \frac{d \tilde V_{\rm LJ}(\tilde r)}{d \tilde r} \, g(\tilde r)~.
}
We checked the pressure obtained through the numerical integration of $g(\tilde r)$ in Eq.~\eqref{eq:Prdf} is consistent with our earlier calculations of the pressure through Eq.~\eqref{eq:pirtheo}.

\section{Comparison with the van der Waals model}
\label{app:vdWcompare}

The van der Waals~(vdW) equation of state is an analytic model for a thermodynamic system of interacting particles exhibiting a first-order phase transition and a CP. 
It has often been used as a simple  model to study the phenomena associated with CP fluctuations~\cite{Vovchenko:2015xja,Vovchenko:2015uda,Poberezhnyuk:2020ayn}.
It is instructive to compare the behavior of particle number fluctuations in this analytic model with the numerical results stemming from microscopic simulations of the LJ fluid.

The scaled variance of particle number fluctuations in the vdW model reads~\cite{Vovchenko:2015xja}
\eq{
\label{eq:wvdW}
\omega_{\rm vdW}(T^*, n^*) = \frac{1}{9}\left[\frac{1}{(3-n^*)^2}-\frac{n^*}{4T^*}\right]^{-1} ~,}
where $n^*= n/ n_{\rm c}$ and $T^* = T/T_{\rm c}$ are the reduced variables normalized to the vdW critical density $n_c$ and temperature $T_c$, respectively. 
To compare the vdW and LJ models one should study a behavior of the scaled variance at the same values of the reduced variables. 
In the LJ model model, these reduced variables are equal to  $T^* = \tilde T/\tilde T_{\rm c}$ and $n^*=\tilde n/\tilde n_{\rm c}$, where $\tilde T_{\rm c}$ and $\tilde n_{\rm c}$ are given by Eq.~(\ref{eq:TcLJ}).

\begin{figure}[!t]
    \includegraphics[width=.48\textwidth]{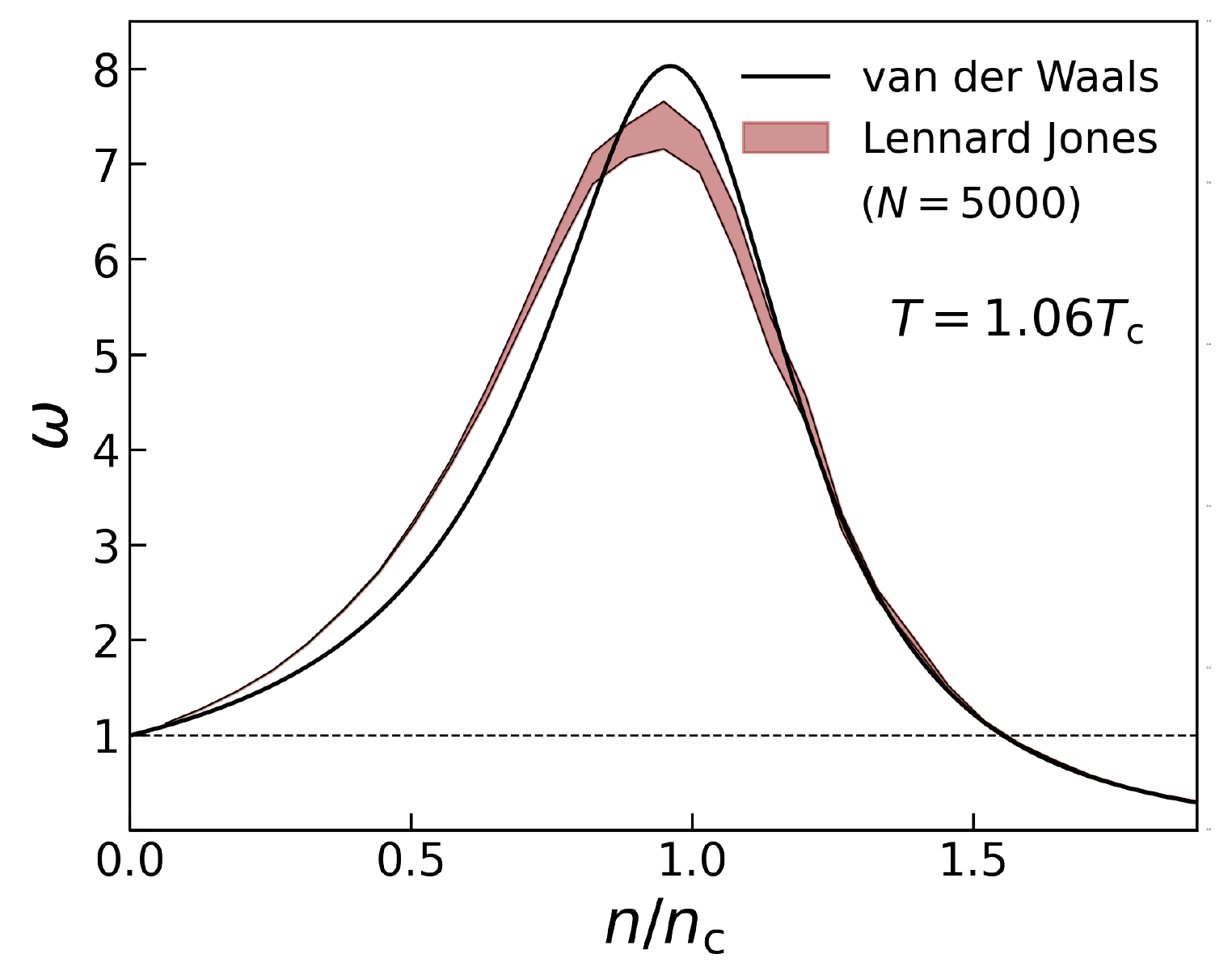}
    \caption{
    The scaled variance of particle number fluctuations in the grand canonical ensemble along the isotherm  $T/T_c=1.06$ as a function of $n/n_c$, calculated analytically via Eq.~\eqref{eq:wvdW} in the van der Waals model~(black line), and numerically via Eq.~\eqref{eq:w} in the Lennard Jones model using molecular dynamics simulations of 5000 particles~(brown band).
    }
    \label{fig:vdw}
\end{figure}

Figure~\ref{fig:vdw} depicts the behavior of the scaled variance $\omega$ at  
$T/T_c =1.06$~($\tilde T  = 1.4$) as a function of $n/n_c$ in the vdW model~(black line) and the LJ model~(brown band).
The vdW model results are analytic~[Eq.~\eqref{eq:wvdW}] while the LJ results correspond to  numerical calculations of $\omega$ via Eq.~\eqref{eq:w} utilizing MD simulations of $N = 5000$ particles. Note that the LJ results here are the same as shown in Fig.~\ref{fig:wi}.

Both models yield qualitatively similar behavior of $\omega$. It grows with $n$  starting from unity at $n \to 0$, exhibits a peak of $\omega \sim 7-8$ near the critical density $n/n_c\sim 1$, and indicates suppressed fluctuations~($\omega < 1$) at larger densities, $n/n_c > 1.5$. Interestingly, the two models show essentially identical results in the density range of $1.2 < n/n_c < 1.9$.
However, a qualitative difference between the two models exists at very high densities.
The maximum achievable densities in the vdW model are restricted by the packing limit at $ n/n_{\rm c}=3$, where fluctuations reach zero, $\omega_{\rm VdW} = 0$.
On the other hand, the LJ model does not contain a hard limit on the maximum achievable densities.

Overall, the presented comparison validates the use of the analytic vdW model for studying qualitative equilibrium features of particle number fluctuations in vicinity of the CP of a first-order phase transition. 
It should be noted, however, that important quantitative differences between the two models do exist. 
For instance, the vdW model corresponds to the mean-field theory universality class, which is different from the Ising universality class characterizing the LJ fluid~\cite{watanabe2012phase}. 
Thus, the two models have different critical scaling laws.

\vskip5pt
\label{app:microcan}









\section{Scaled variance in the microcanonical ensemble}
\label{app:microcan}

Here we calculate the scaled variance for a sub-system where the energy
of the total system in conserved. In other words the total system
is governed by a microcanonical ensemble. We consider a system
of non-interacting and non-relativistic particles. The subsystem is
defined by considering only particles with a z-component of their
momenta to be within the acceptance region, $\left|p_{i,z}\right|<p_z^{\rm {cut}}$~(or equivalently $|v_{i,z}| < v_z^{\rm cut}$ given that $p_{i,z} = m v_{i,z}$).

\begin{widetext}

Let us start with the microcanonical partition function for a system
with energy $E$ containing $N$ non-interacting particles of mass $m$.
This system has $k=3N$ degrees of freedom and its partition function
is related to the surface $S_{k}(R)$ of a sphere in $k$ dimensions:
\[
Z=A\int dp_{1}\ldots dp_{k}\,\delta\left(2mE-\sum_{i=1}^{k}p_{i}^{2}\right)=\frac{A}{2R}\int dp_{1}\ldots dp_{k}\delta\left(R-\sqrt{\sum_{i=1}^{k}p_{i}^{2}}\right)=A\frac{S_{k}(R)}{2R}
\]
Here $R=\sqrt{2mE}$ and $A$ is an irrelevant constant. Since the acceptance cuts only
affect the $z$-components of the momenta, we subsequently denote by
$p_{i}$ the $z$-component of particle $i$.
The probability to find
a particle with $z$-momentum $p_{1}$, $w_{1}\left(p_{1}\right)$, is
then given by
\[
w_{1}\left(p_{1}\right)=\frac{A}{Z}\int dp_{2}\ldots dp_{k}\,\delta\left[\left(2mE-p_{1}^{2}\right)-\sum_{i=2}^{k}p_{i}^{2}\right]=\frac{R}{\sqrt{R_{1}^{2}-p_{1}^{2}}}\frac{S_{k-1}\left(\sqrt{R_{1}^{2}-p_{1}^{2}}\right)}{S_{k}(R)}.
\]
Similarly, the probability $w_{2}\left(p_{1},p_{2}\right)$ to find
a pair of particles with z-momenta $p_{1}$ and $p_{2}$ is given
by
\[
w_{2}\left(p_{1},p_{2}\right)=\frac{A}{Z}\int dp_{3}\ldots dp_{k}\,\delta\left[\left(2mE-p_{1}^{2}-p_{2}^{2}\right)-\sum_{i=3}^{k}p_{i}^{2}\right]=\frac{R}{\sqrt{R_{1}^{2}-p_{1}^{2}-p_{2}^{2}}}\frac{S_{k-2}\left(\sqrt{R_{1}^{2}-p_{1}^{2}-p_{2}^{2}}\right)}{S_{k}(R)}.
\]
Using the well-known formula for the surface of an $k$-dimensional sphere
\[
S_{k}(R)=2\frac{\pi^{k/2}}{\Gamma\left(\frac{k}{2}\right)}R^{k-1}
\]
one finds for the single particle probability 
\[
w_{1}\left(p_{1}\right)=\frac{dN}{dp_{1}}=\frac{R^{2-k}\left(R^{2}-p_{1}^{2}\right)^{\frac{k-3}{2}}\Gamma\left(\frac{k}{2}\right)}{\sqrt{\pi}\Gamma\left(\frac{k-1}{2}\right)}
\]
and for the two particle probability 
\begin{align*}
w_{2}\left(p_{1},p_{2}\right) & =\frac{d^{2}N}{dp_{1}dp_{2}}=\frac{R^{2-k}\left(R^{2}-p_{1}^{2}-p_{2}^{2}\right)^{\frac{k-4}{2}}\,\left(k-2\right)}{2\pi}
\end{align*}
Given the one particle and two-particle probabilities, $w_{1}\left(p_{1}\right)$
and $w_{2}\left(p_{1},p_{2}\right)$, the mean number $\ave n$ and
number of pairs, $\ave{n(n-1)}$ for the acceptance region are
\begin{align}
\ave n & =N\int_{-p_{\rm cut}}^{p_{\rm cut}}w_{1}(p_{1})dp_{1}\label{eq:app_mean}\\
\ave{n(n-1)} & =N(N-1)\int_{-p_{\rm cut}}^{p_{\rm cut}}dp_{1}\int_{-p_{\rm cut}}^{p_{\rm cut}}dp_{2}\,w_{2}\left(p_{1},p_{2}\right).\label{eq:app_pairs}
\end{align}
Here $p_{\rm cut} \equiv p_z^{\rm cut}$ and $N$ denotes the (conserved) total number of particles in the
entire system. The variance $\text{var}(n)=\ave{n^{2}}-\ave n^{2}$ is easily
obtained 
\[
\text{var}(n)=\ave{\left(\delta n\right)^{2}}=\ave{n(n-1)}+\ave n-\ave n^{2}.
\]
The integrals in Eqs. \eqref{eq:app_mean} and \eqref{eq:app_pairs} can
be evaluated with, for example Mathematica, analytically for Eq. \eqref{eq:app_mean},
and numerically for Eq. \eqref{eq:app_pairs}. Here, we are interested
in the limit of large number of degrees, i.e. $k\rightarrow\infty$.
To this end it is convenient to introduce scaled and dimensionless
momenta
\[
q=\frac{\sqrt{k}}{R}p
\]
This choice of variable is motivated by the canonical limit in which
case $w_{1}(p_{1})\sim\exp\left[\frac{p_{1}^{2}}{2mT}\right]=\exp\left[\frac{p_{1}^{2}}{2(2mE)}k\right]=\exp\left[\frac{q_{1}^{2}}{2}\right]$.
This scaling removes the trivial dependence of the typical momentum
on the number of degrees of freedom and ensures that most of the particles
will have a rescaled momentum of $\left|q\right|\lesssim1$. The single
and two-particle probabilities are then
\begin{align*}
w_{1}\left(q_{1}\right) & =\frac{dN}{dq_{1}}=\frac{dp_{1}}{dq_{1}}\frac{dN}{dp}=\frac{R}{\sqrt{k}}\frac{dN}{dp}=\frac{k^{\frac{2-k}{2}}\left(k-q_{1}^{2}\right)^{\frac{k-3}{2}}\Gamma\left(\frac{k}{2}\right)}{\sqrt{\pi}\Gamma\left(\frac{k-1}{2}\right)}\\
w_{2}\left(q_{1},q_{2}\right) & =\frac{d^{2}N}{dq_{1}dq_{2}}=\frac{R^{2}}{k}\frac{d^{2}N}{dp_{1}dp_{2}}=\frac{(k-2)\,k^{\frac{2-k}{2}}\left(k-q_{1}^{2}-q_{2}^{2}\right){}^{\frac{k-4}{2}}}{2\pi}.
\end{align*}

\end{widetext}

\begin{figure}[!t]
\centering
\vskip10pt
\includegraphics[width=.45\textwidth]{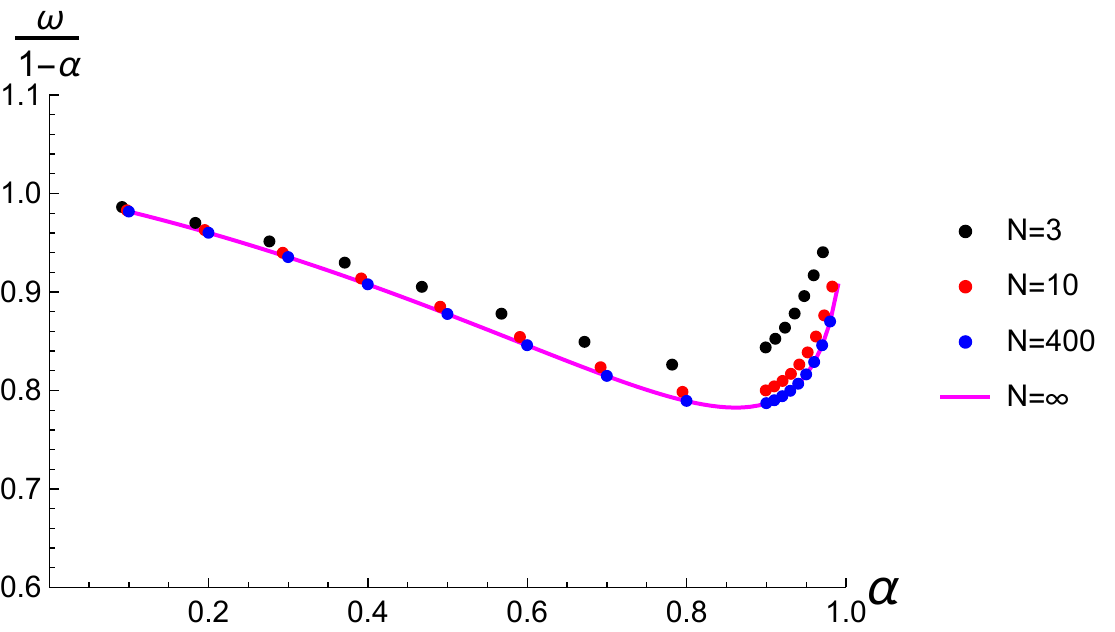} 
\caption{Scaled variance versus the fraction of particles in acceptance for
$N=$3, 10, and 400 is depicted by black~(upper), red~(middle), and blue~(lower) circles. Also shown is the analytic result
for the limit of $N\rightarrow\infty$ (solid magenta line).}
\label{fig:microcomp}
\end{figure}

The mean number $\ave n$ and number of pairs, $\ave{n(n-1)}$ are
then given by integrals of the scaled momenta $q$ similar to expressions \eqref{eq:app_mean} and \eqref{eq:app_pairs} but with scaled
integration limits, $q_{cut}=\frac{\sqrt{k}}{R}p_{cut}$. The limit
for large number of degrees of freedom is obtained by first expanding
$w_{1}\left(q_{1}\right)$ and $w_{2}\left(q_{1},q_{2}\right)$ in
powers of $1/k$ and then integrating over the acceptance interval in
order to obtain the mean number and number of pairs. Finally one takes
the limit of $k\rightarrow\infty$ keeping in mind that both the mean
and the variance should scale with the number of degrees of freedom.
As a result one obtains
\begin{align*}
\ave n & \underset{k\rightarrow\infty}{\rightarrow}\frac{1}{3}k\,\text{erf}\left(\frac{q_{\text{cut}}}{\sqrt{2}}\right)\\
\ave{\left(\delta n\right)^{2}} & \underset{k\rightarrow\infty}{\rightarrow}\frac{1}{9}k\left[3\text{\,erf}\left(\frac{q_{\text{cut}}}{\sqrt{2}}\right)\text{erfc}\left(\frac{q_{\text{cut}}}{\sqrt{2}}\right)-\frac{e^{-q_{\text{cut}}^{2}}q_{\text{cut}}^{2}}{\pi}\right].
\end{align*}
Since $N=3k$ the fraction of particles $\alpha=\frac{\ave n}{N}$
is then given by 
\begin{equation}
\alpha=\text{erf}\left(\frac{q_{\text{cut}}}{\sqrt{2}}\right)\label{eq:alpha}
\end{equation}
Consequently one obtains the following for the scaled variance divided by the charge conservation correction, $\tilde{\omega}^{\rm mom,mce}_{\rm id} = \omega / (1-\alpha)$:
\[
\tilde{\omega}^{\rm mom,mce}_{\rm id} = 1-\frac{e^{-q_{\text{cut}}^{2}}q_{\text{cut}}^{2}}{3\pi\left[\text{erf}\left(\frac{q_{\text{cut}}}{\sqrt{2}}\right)\text{erfc}\left(\frac{q_{\text{cut}}}{\sqrt{2}}\right)\right]}
\]

Eq. \eqref{eq:alpha} allows to express the the cutoff momentum $q_{\text{cut}}$
in terms of the fraction of accepted particles, $q_{\text{cut }}=\sqrt{2}\:\text{erf}^{-1}(\alpha)$
so that
\begin{equation}
\tilde{\omega}^{\rm mom,mce}_{\rm id} = 1-\frac{2e^{-2\text{erf}^{-1}(\alpha)^{2}}\text{erf}^{-1}(\alpha)^{2}}{3\pi\alpha\left(1-\alpha\right)}.\label{eq:large_k}
\end{equation}

In Fig.~\ref{fig:microcomp} we show the comparison of the the above result, Eq. \eqref{eq:large_k}, 
labeled as $N=\infty$, together with the explicit the results obtain
by numerically integrating Eqs.~\eqref{eq:app_mean} and \eqref{eq:app_pairs}
for $N=3,\,10,\,\text{and}\,400$. 
One can see that Eq.~\eqref{eq:large_k} describes the fluctuations qualitatively even in small systems~($N = 3$ and $10$), while for $N = 400$ or higher the description is very accurate quantitatively.

We also verified the analytic results by performing Monte Carlo of particle momenta with the constraint of exact total energy conservation. The Monte Carlo results are in good agreement with the analytic results.
Using Monte Carlo we also checked the additional effect of exact conservation of total momentum in addition to energy conservation, and this effect turned out to have a negligible influence on the behavior of $\tilde{\omega}^{\rm mom,mce}_{\rm id}$.


\bibliography{references.bib}
\end{document}